\documentclass{iaus}
\usepackage{graphicx}

\title[The halo radial temperature structure in NGC7662] 
{New observations of the halo radial temperature structure in NGC7662}
\author[C.\ Sandin, D.\ Sch\"onberner, M.M.\ Roth et al.]   
{C.\ Sandin, D.\ Sch\"onberner, M.M.\ Roth, M.\ Steffen, \break A.\ Monreal-Ibero, P.\ B\"ohm, U.\ Tripphahn}

\affiliation{Astrophysikalisches Institut Potsdam, An der Sternwarte 16, D-14482 Potsdam, Germany \break
email: CSandin@aip.de}

\pubyear{2006}
\volume{234}
\pagerange{xxx--xxx}
\date{03/05/2006 and in revised form 03/05/2006}
\jname{Planetary Nebulae in our Galaxy and Beyond}
\editors{M. J. Barlow \& R. H. M\'endez, eds.}

\def\arcsec{\ensuremath{^{\prime\prime}}}
\begin{document}

\maketitle

\begin{abstract}
We report on our studies of the physical structure of the planetary nebula (PN) NGC7662. Using (3D) Integral Field Spectroscopy we have been able to measure the electron temperature more accurately and at a larger number of radial locations than before. Here we briefly present our method by which we find a strong positive temperature gradient with increasing radius. According to hydrodynamic models a hot halo, when compared to the central star, can be the product of the passage of an ionization front (e.g.\ Marten 1993). Such a gradient is not found in equilibrium models, and this finding -- when confirmed for other objects -- strongly advocates the use of hydrodynamic models when modeling PN halos.
\end{abstract}

\firstsection 
\section{Observations}
The data presented here was obtained with the PMAS instrument placed on the 3.5m telescope at Calar Alto in September 2005. In three different pointings, forming a \emph{mosaic}, the PMAS integral field unit (IFU) covered a distance out to 40{\arcsec} west of the central star. The wavelength coverage was 3436--5093\,{\AA} with a spectral resolution of 3.6\,{\AA}. The IFU holds $16\times16$ separate fibers that each represents a so called \emph{spaxel} -- covering 1{\arcsec}$\times$1{\arcsec} on the sky. Any number of spaxels can be combined to create a final spectrum with an increased S/N ratio, compared to that of a single spaxel. We have extracted six bands of spaxels from the three pointings. The flux in the spectra of the consecutive bands at increasing radius is a sum of (the spatial width in arcseconds of each band is specified as a subscript): $2_{1}$, $4_{1}$, $16_{2}$, $36_{3}$, $63_{4}$, and $128_{8}$ spaxels, respectively.

\section{Results}
To calculate the electron temperature $T_\mathrm{e}$ at each of the six radial positions represented by the bands we use the three [O{\sc iii}] 4363, 4959, 5007\,{\AA} lines out of which [O{\sc iii}] 4363\,{\AA} is the weakest. Since we do not have a suitable sky frame the flux of the Hg\,4359\,{\AA} sky line is removed from that of [O{\sc iii}] 4363\,{\AA} by decomposition. We note that \cite{Middlemass91} in a previous observation of NGC7662 removed the same sky line using a sky frame in which a significant amount of flux appears to be present in the [O{\sc iii}] line (see Fig.\ 3b ibid.). This most likely indicates that the halo extends to radii larger than where the sky was measured. By accounting for this ``missing'' flux we have found that their measured [O{\sc iii}] temperature increases by a factor of about 2.5 (see below).

\begin{figure}[t]
\begin{minipage}[t]{6.5cm}
\centerline{\includegraphics{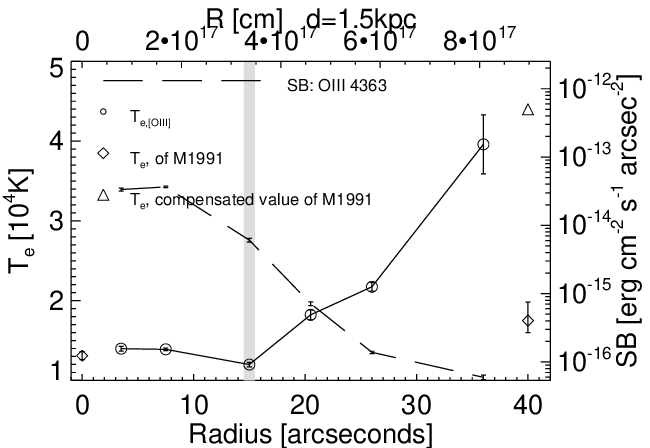}}
\caption{Radial temperature structure}
\end{minipage}
\hfill
\begin{minipage}[t]{6.5cm}
\centerline{\scalebox{0.40}{%
\includegraphics[width=18cm,clip,bb=14 28 600 415]{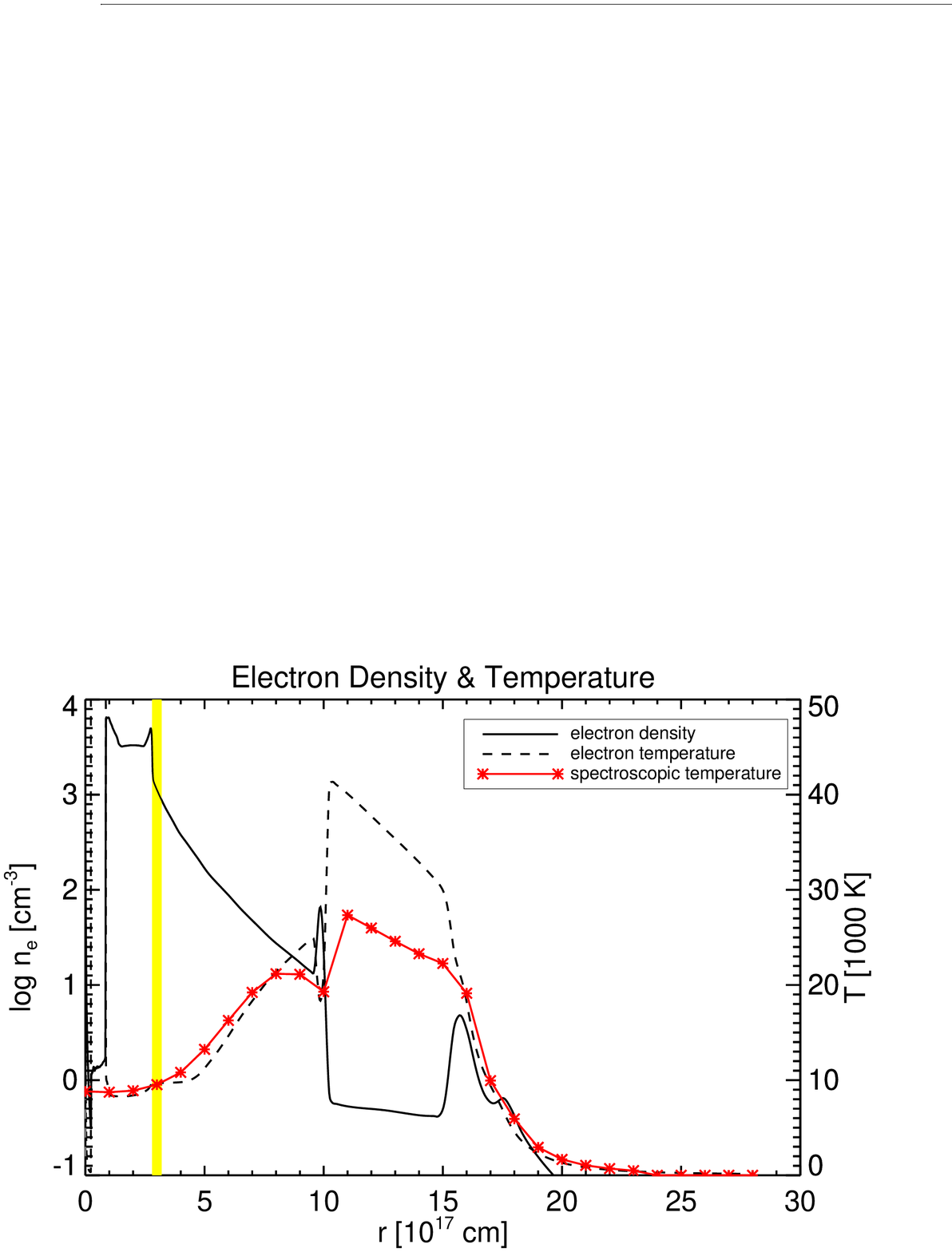}%
}}
\caption{Hydrodynamic PN model}
\end{minipage}
\end{figure}

The radial temperature structure is presented together with the surface brightness (SB) of [O{\sc iii}] 4363\,{\AA} and error estimates in Fig.~1. A strong temperature gradient is seen in the temperature outside the PN in the halo at radii $\ge15\arcsec$ -- the border between the central parts of the PN and the halo is marked with a gray bar. A possible theoretical explanation for such a strong gradient in the halo is given in Sect.~\ref{sec:theory}. Also note that the temperature measured at $R=36\arcsec$ is an average over $8\arcsec$ why measurements of increased resolution will yield even higher temperatures in this region of the halo where the gradient is the strongest.
For comparison the two corresponding temperatures sampled by \cite{Middlemass91} are also shown (M1991). According to the discussion above we have calculated a new value for their estimate at $R=40\arcsec$. The resulting scaled temperature $T_{\mathrm{e}}=44\,000$\,K is in better agreement with our calculations than the previous value of $T_{\mathrm{e}}=17\,500$\,K.

\section{Theoretical modeling}\label{sec:theory}
In Fig.~2 we show a hydrodynamic PN model shortly after the passage of an ionization front. The density trough at $1.0\cdot10^{18}\!\le\!R\!\le\!1.6\cdot10^{18}$\,cm is due to the final thermal pulse on the AGB. The line marked with ``*'' is the [O{\sc iii}] temperature as one would get by applying standard plasma diagnostics; it shows a positive gradient out to $R=11\cdot10^{17}$cm, similar to the ``observed'' structure in Fig.~1. The gradient is due to non-equilibrium heating and cooling immediately after the passage of the ionization front (cf.\ Marten 1993 and Tylenda 2003). Note that this model has not been chosen to fit observed properties. An even steeper temperature gradient could be achieved if the density was lower in the inner parts
since modeled peak temperatures depend on the density (radiative cooling is $\propto\rho^2$).

\section{Future work}\label{sec:fut}
As a continuation to the work presented here we will in the next step, using observations and appropriate models, study the physical structure of several additional PNe that are found to have a halo.
Finally, new automated software will allow a more precise determination of halo PN information found in spectra of integral field units than before.

\begin{acknowledgments}
C.S.\ is supported by DFG under grant number SCHO 394/26.
\end{acknowledgments}

\end{document}